\documentclass[aps,prd,showpacs,twocolumn,floatfix,nofootinbib,superscriptaddress,
amsmath,amssymb,amsfonts,]{revtex4-1} 

\usepackage{bm}
\usepackage{slashed}
\usepackage{graphicx}

\usepackage{amsmath,amssymb,amsfonts,graphicx}
\usepackage{dsfont, mathrsfs}
\usepackage{bm}
\usepackage{slashed}
\usepackage{subfigure}
\usepackage[svgnames]{xcolor}
\usepackage{placeins}
\usepackage{array}
\usepackage{natbib,hyperref}  
\usepackage{url}
\usepackage[utf8]{inputenc}

\makeatletter
\newcommand{\thickhline}{%
    \noalign {\ifnum 0=`}\fi \hrule height 1pt
    \futurelet \reserved@a \@xhline
}
\newcolumntype{"}{@{\hskip\tabcolsep\vrule width 1pt\hskip\tabcolsep}}
\makeatother

\usepackage{placeins}

\usepackage{xspace}

\newcolumntype{L}[1]{>{\raggedright\let\newline\\\arraybackslash\hspace{0pt}}m{#1}}
\newcolumntype{C}[1]{>{\centering\let\newline\\\arraybackslash\hspace{0pt}}m{#1}}
\newcolumntype{R}[1]{>{\raggedleft\let\newline\\\arraybackslash\hspace{0pt}}m{#1}}
\newcommand{\Eq}[1]{Eq.~(\ref{#1})}
\newcommand{\al}[1]{\begin{align} #1 \end{align}}
\newcommand{\non}{\nonumber}
\newcommand{\vect}[1]{\boldsymbol{#1}}

\def\fh{\varphi_h}

\def\fR{\varphi_R}

\def\fK{\varphi_{k}}

\def\fRK{\varphi_{RK}}
\def\kT{\vect{k}_T}

\def\RT{\vect{R}_T}

\def\psq{p_\perp^2}
\newcommand{\psqn}[1]{p_{#1\perp}^2}

\def\PP{\vect{P}_\perp}
\newcommand{\PPn}[1]{\vect{P}_{#1\perp}}

\newcommand{\pe}[1]{\vect{p}_{{#1}\perp}}
\newcommand{\Pe}[1]{\vect{P}_{{#1}\perp}}

\def\sT{\vect{s}_T}

\newcommand{\sq}[1]{\vect{s}_{#1}}

\newcommand{\ImL}{0.8\columnwidth}
\newcommand{\GapCapt}{\vspace{-8pt}}
\newcommand{\GapSubf}{\vspace{-8pt}}
\begin{document}

\title{Dihadron fragmentation functions in the quark-jet model:\\ Longitudinally polarized quarks}

\preprint{ADP-17-30/T1036}

\author{Hrayr~H.~Matevosyan}
\thanks{ORCID: http://orcid.org/0000-0002-4074-7411}
\affiliation{ARC Centre of Excellence for Particle Physics at the Tera-scale,\\ 
and CSSM, Department of Physics, \\
The University of Adelaide, Adelaide SA 5005, Australia
\\ http://www.physics.adelaide.edu.au/cssm
}

\author{Aram~Kotzinian}
\thanks{ORCID: http://orcid.org/0000-0001-8326-3284}
\affiliation{Yerevan Physics Institute,
2 Alikhanyan Brothers St.,
375036 Yerevan, Armenia
}
\affiliation{INFN, Sezione di Torino, 10125 Torino, Italy
}

\author{Anthony~W.~Thomas}
\thanks{ORCID: http://orcid.org/0000-0003-0026-499X}
\affiliation{ARC Centre of Excellence for Particle Physics at the Tera-scale,\\     
and CSSM, Department of Physics, \\
The University of Adelaide, Adelaide SA 5005, Australia
\\ http://www.physics.adelaide.edu.au/cssm
}

\begin{abstract}
We study the dihadron fragmentation functions (DiFF) of longitudinally polarized quarks into pion pairs. The quark-jet framework is used to model the sequential hadronization of a polarized quark into hadrons, where the polarization transfer to the remnant quark in each hadron emission step is calculated using the spin density matrix formalism. Using Monte Carlo (MC) simulations of the hadronization process, we find a nonvanishing helicity dependent DiFF,  $G_1^\perp$ , which is also related to the longitudinal jet handedness. A method is developed for extracting the angular moments of this DiFF, which enter the expressions for the azimuthal asymmetries for an electron-positron annihilation process into two pairs of hadrons from back-to-back  jets and the dihadron production in semi-inclusive deep inelastic scattering. Finally, we derive explicit integral expressions for DiFFs where only two hadrons are emitted by a quark and use them to validate our MC results for both unpolarized and helicity dependent DiFFs.
\end{abstract}



\date{\today}                                           

\maketitle

\section{Introduction}
\label{SEC_INTRO}
 
In recent years a great deal of attention has been paid to the study of the spin structure of the nucleon using deep inelastic, semi-inclusive hadron production (SIDIS) with two measured, unpolarized final state hadrons~\cite{Bacchetta:2011ip, Bacchetta:2012ty, Pisano:2015wnq}. It was shown in Refs.~\cite{Bianconi:1999cd, Bianconi:1999uc, Radici:2001na,Bacchetta:2003vn} that this process allows one to directly extract the collinear transversity parton distribution functions (PDF) from a specific azimuthal single spin asymmetry (SSA), where it is multiplied by the so-called interference DiFF (IFF) $H^\sphericalangle_1$. At the same time, the two-hadron SIDIS cross section on a longitudinally polarized nucleon also gives access to the helicity PDF, which in this case is convoluted with the helicity dependent DiFF $G_1^\perp$. The latter is interesting both because it has no analogue in the single unpolarized hadron production case and because it is related to the long-predicted quantity of longitudinal jet handedness~\cite{Efremov:1992pe}. Both $H^\sphericalangle_1$ and  $G_1^\perp$ are T-odd and can be extracted from the azimuthal modulations in two hadron pair production process from back to back jets in $e^+e^-$ annihilation~\cite{Boer:2003ya}. The measurements from the BELLE experiment~\cite{Vossen:2011fk} yielded significant results for the asymmetry involving IFF $H_1^\sphericalangle$, that  were used in Refs.~\cite{Courtoy:2012ry,Radici:2015mwa} to fit  parametrizations of $H_1^\sphericalangle$. Recently the BELLE experiment produced preliminary results on SSA sensitive to $G_1^\perp$, showing no signal within the experimental  uncertainties~\cite{Abdesselam:2015nxn, Vossen:2015znm}.  Similarly, the recent results from the COMPASS experiment~\cite{Sirtl:2017rhi} also showed no significant signal for SIDIS asymmetry that involve helicity dependent DiFF.

 The DiFFs have been already studied in the quark-jet model, both for the case of an unpolarized~\cite{Matevosyan:2013aka, Matevosyan:2013nla} and transversely polarized quark~\cite{Matevosyan:2013eia}. In the first studies the simplistic treatment of the polarization transfer during the quark hadronization nevertheless created unphysical modulations that had to be circumvented using additional assumptions. However, more recently the quark-jet hadronization framework has been developed to self-consistently describe the hadronization of a quark~\cite{Bentz:2016rav} with any polarization, where the polarization transfer from the fragmenting to the remnant quark in each hadron emission step has been calculated using the complete set of twist-two transverse momentum dependent (TMD) quark-to-quark splitting functions. Moreover, MC implementation of the polarized quark hadronization based on this framework was developed and implemented in Ref.~\cite{Matevosyan:2016fwi}, and both the unpolarized and the Collins fragmentation functions of pions produced in the hadronization of a light quark were computed. The input elementary quark-to-quark splitting functions (SF) used in that work were derived using Nambu--Jona-Lasinio (NJL) effective theory of quark interactions~\cite{Nambu:1961tp,Nambu:1961fr}. 
 
 In this work, we use this MC implementation of the quark-jet model to study the dihadron correlations in the hadronization of a quark with nonzero longitudinal polarization. We aim to calculate the relative size of the helicity dependent DiFF compared to the unpolarized DiFF for pion pairs. Further, we derive integral expressions for both these DiFFs in the case where only two hadrons are emitted by the quark in order to validate our MC results using an alternate method. Finally, these derived integral expressions will elucidate the mechanism for generating the dihadron correlations encoded by $G_1^\perp$. These results are independent of the transverse polarization component of the quark.

 This paper is organized in the following way.  In the next section we briefly review the formalism for DiFFs. In Sec.~\ref{SEC_MC} we briefly describe the details of the MC simulations and the newly developed method for extracting $G_1^\perp$. We also present the results for the simulations with different numbers of hadrons produced by the fragmenting quark. In Sec.~\ref{SEC_VALIDATION} we validate the MC results against integral expressions for the unpolarized and helicity dependent DiFFs derived for two-step hadronization process.  In that section we also discuss the dynamical mechanism for generating helicity dependent two-hadron correlations. We present our conclusions in Sec.~\ref{SEC_CONCLUSIONS}.

\section{Formalism of the DiFFs}
\label{SEC_DIFF_FORM}
 
  In this section we briefly review the kinematics and the formal definitions of the DiFFs, following Refs.~\cite{Bianconi:1999cd, Bianconi:1999uc, Radici:2001na,Bacchetta:2003vn, Boer:2003ya}. The kinematics of dihadron fragmentation is described by the momentum $k$ of the initial quark $q$ with mass $m$ fragmenting into two hadrons $h_1,h_2$ with momenta $P_1, P_2$ and masses $M_1,M_2$. We define the light-cone components of a 4-vector $a$ as $a^\pm = \frac{1}{\sqrt{2}}(a^0 \pm a^3)$. The light-cone momentum fractions of the two hadrons are then defined as $z_i = {P_i^+}/{k^+}$. The total $P$ and the relative $R$ momenta are defined as
\al
{
&
 P \equiv P_h = P_1 + P_2,
 \\
 &
 R = \frac{1}{2}( P_1 - P_2),
 }
 and the relevant light-cone momentum fractions are
\al
{
 z &= z_1  + z_2,
\\
 \xi &= \frac{z_1}{z}  = 1- \frac{z_2}{z} \, .
}

\begin{figure}[t]
\centering 
\subfigure[] {
\includegraphics[width=\ImL]{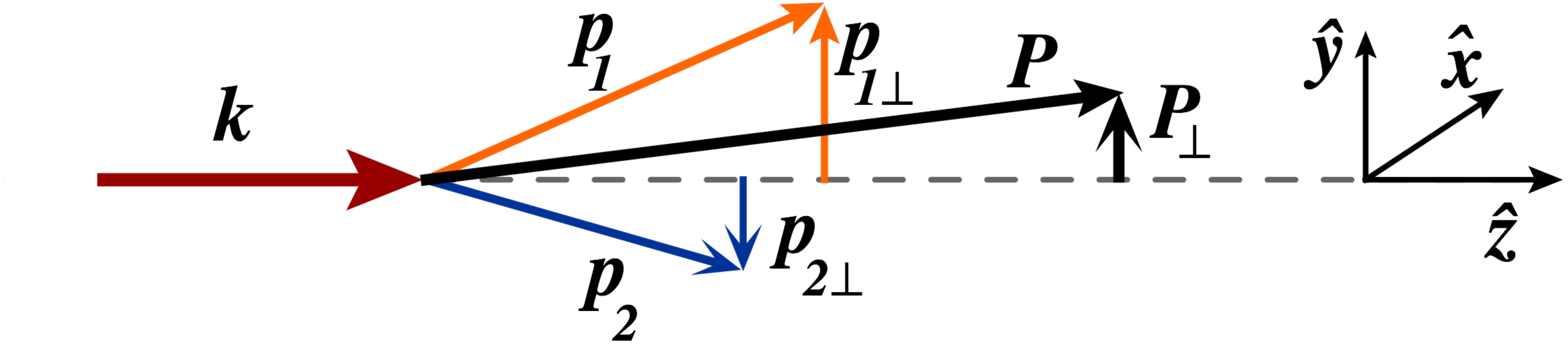}
}
\\ 
\subfigure[] {
\includegraphics[width=\ImL]{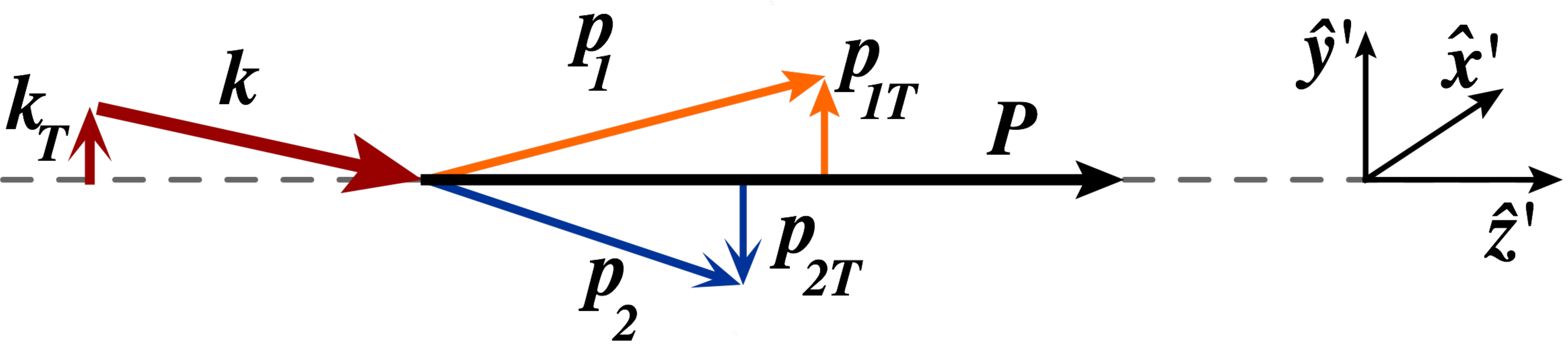}
}
\GapCapt
\caption{Common frames used in defining DiFFs: (a) the $\perp$ system  where the fragmenting quark's 3-momentum $\vect{k}$ is along the $\hat{z}$ axis  (b) the $T$ system where the total 3-momentum of hadrons $\vect{P}$ is along the $\hat{z}'$ axis.}
\label{PLOT_PERP_T_SYS}
\vspace{1cm}
\end{figure}

The two most common coordinate systems used in describing DiFFs are the so-called $\perp$ and $T$ systems depicted in Figs.~\ref{PLOT_PERP_T_SYS} (a) and (b), respectively. We denote the components of a 3-vector $\vect{a}$, perpendicular to the $\hat{z}$ and $\hat{z}'$ axes in these two systems, as $\vect{a}_\perp$ and $\vect{a}_T$.

We can relate the $\perp$ and $T$ components of vectors in the two systems by considering a Lorentz transformation, which preserves the light-cone momentum fractions and results in quark acquiring transverse momentum $\kT$ in the $T$ system. The corresponding transformation matrix reads~\cite{Diehl:2000xz}
\al
{
\Lambda^\mu_\nu=
 \left(
 \begin{array}{cccc}
 1 & \frac{\kT^2}{(k^+)^2} & \frac{k^1}{k^+} & \frac{k^2}{k^+} \\
 0 & 1 & 0 & 0 \\
 0 & \frac{k^1}{k^+} & 1 & 0 \\
 0 & \frac{k^2}{k^+} & 0 & 1
 \end{array}
 \right),
 \label{EQ_LORENTZ}
}
where $\mu,\nu \in \{+, -, 1, 2\}$. Then one easily finds
\al
{
 \vect{P}_{1T} &= \PPn{1}  + z_1 \kT,
  \\
 \vect{P}_{2T} &= \PPn{2}  + z_2 \kT \, .
 }
 The relations for the total and relative transverse momenta of the hadron pair imply that:
 \al{
 &
\kT = -\frac{\PP}{z},
 \\&
 \RT = \frac{z_2 \PPn{1} - z_1 \PPn{2}}{z} = (1-\xi) \PPn{1} - \xi \PPn{2}.
}

The formal definition of DiFFS is given using the so-called quark-quark correlator in the $T$ system~\cite{Bianconi:1999cd, Radici:2001na, Boer:2003ya}
\al
{
 \Delta_{ij}(k; P_1, P_2)& 
 \\ \non
 = \sum_X \int d^4 \zeta &e^{i k\cdot \zeta} \langle 0 |\psi_i(\zeta) | P_1 P_2, X \rangle \langle P_1 P_2, X | \bar{\psi}_j(0) | 0 \rangle.
 }
In particular, the DiFFs are defined via projections of the quark-quark correlator, defined for a Dirac operator $\Gamma$ as
\al
{
 \Delta^{\Gamma}&(z, \xi, \kT^2, \RT^2, \kT \cdot \RT) 
 \\ \non
 &= \frac{1}{4z} \int d k^+ \mathrm{Tr}[\Gamma \Delta(k, P_1, P_2)]|_{k^+ = P_h^+/z}.
}
The expressions for all four leading order DiFFs of a polarized quark into an unpolarized hadron pair are
\al
{
\label{EQ_DELTA_UNP}
 \Delta^{[\gamma^+]} =& D_1(z, \xi, \kT^2, \RT^2, \kT \cdot \RT),
\\ \label{EQ_DELTA_LIN}
  \Delta^{[\gamma^+\gamma_5]} 
  =& \frac{\epsilon_T^{ij} R_{Ti} k_{Tj}}{M_1 M_2} G_1^\perp(z, \xi, \kT^2, \RT^2, \kT \cdot \RT),
 \\ \label{EQ_DELTA_TRANSV}
 \Delta^{[i \sigma^{i+} \gamma_5]} 
  =& \frac{\epsilon_T^{ij} R_{Tj}}{M_1 + M_2} H_1^\sphericalangle(z, \xi, \kT^2, \RT^2, \kT \cdot \RT) 
  \\ \non
& + \frac{\epsilon_T^{ij}  k_{Tj}}{M_1 + M_2} H_1^\perp(z, \xi, \kT^2, \RT^2, \kT \cdot \RT),
}
where $D_1$ is the unpolarized DiFF and $G_1^\perp$ the helicity dependent DiFF describing the correlations between the total and relative transverse momenta of the pair and the longitudinal polarization of the quark. The two remaining DiFFs describe the correlations between the transverse polarization of the quark and the transverse momenta: the analogue of the Collins function $H_1^\perp$ encodes the correlations with the total transverse momentum, while the  IFF $H_1^\sphericalangle$ encodes correlations with the relative transverse momentum.

 The relevant parts of the integrated cross section for back-to-back creation of two dihadron pairs in $e^+e^-$ annihilation involve only the following integrals of the DiFFs~\cite{Boer:2003ya}
\al
{
 \label{EQ_D1_MH}
 D_1(z, M_h^2)
 = &  \int d \xi  \int d \fR  \int d^2 \kT
 \\ \non 
 & \times \ D_1(z, \xi, \kT^2, \RT^2, \kT \cdot \RT),
 }
\al
{
 \label{EQ_G1_MH}
 G_1^\perp(z, M_h^2)
  =&  \int d \xi  \int d \fR  \int d^2 \kT  
\\ \non
 & \times (\kT\cdot \RT) \ G_1^\perp(z, \xi, \kT^2, \RT^2, \kT \cdot \RT) \, 
}
while for the transverse polarization-dependent DiFFs
\al
{
 \label{EQ_HAng_MH}
 H_1^\sphericalangle(z, M_h^2)
  = & \int d \xi  \int d \fR  \int d^2 \kT 
 \\ \non 
 & \times   |\RT| H_1^\sphericalangle(z, \xi, \kT^2, \RT^2, \kT \cdot \RT),
}
\al
{
 \label{EQ_HPerp_MH}
 H_1^\perp(z, M_h^2)
 = &   \int d \xi \int d \fR  \int d^2 \kT
 \\ \non
 & \times  |\kT|  \ H_1^\perp(z, \xi, \kT^2, \RT^2, \kT \cdot \RT),
}
where $\fR$ is the azimuthal angle of vector $\RT$ and the invariant mass of the hadron pair $M_h$ is employed to replace $|\RT|$ in the integrand using the relation
\al
{
R_T^2  &=  \xi (1-\xi) M_h^2 - M_1^2 (1-\xi) - M_2^2 \xi \, ,
\\
M_h^2 &= P^2.
}
 
 The unintegrated DiFFs in Eqs.~(\ref{EQ_DELTA_UNP})-(\ref{EQ_DELTA_TRANSV}) are even functions of the relative azimuthal angle $\fRK \equiv \fR-\fK$ between the vectors $\kT$ and $\RT$, since they only depend on the scalar product $(\kT \cdot \RT)$. Thus, in general their Fourier decomposition in angle $\fRK$ would involve only cosine moments. We define the $n$-th Fourier cosine moment of $G_1^\perp$ in expansion withe respect to $\fRK$ as
\al
{
 \label{EQ_G1_MOM_MH}
 &G_1^{\perp,[n]}(z, M_h^2)
  =  \int d \xi  \int d \fR  \int d^2 \kT  
\\ \non
 & \times  |\kT|   |\RT| \cos(n \cdot \fRK) \ G_1^\perp(z, \xi, \kT^2, \RT^2, \kT \cdot \RT).
}

We note that in  $e^+e^-$ annihilation cross section contains only the first moment of this decomposition, shown in \Eq{EQ_G1_MH}. On the other hand, the $F_{UL}^{\sin(\fh-\fR)}$ structure function (see, e.g. Eq.~(B3) in Ref.~\cite{Bianconi:1999cd}) in SIDIS cross section  can in general be decomposed into infinite Fourier series with respect to $\cos(\fh-\fR)$. In turn, these moments can be expressed as  convolutions of helicity PDF and combinations of Fourier cosine moments of $G_1^{\perp,[n]}$. A similar approach using decomposition in terms of spherical harmonic with azimuthal angle $\fRK$ has been presented in Ref.~\cite{Gliske:2014wba}. Naturally, the Fourier cosine moments we use here can be related to these spherical harmonics, as they both encode the same initial functions.

%
\section{Monte Carlo simulations}
\label{SEC_MC}

\subsection{Extracting DiFFs from  polarization-dependent number densities}
\label{SUBSEC_DIFF_EXTRACT}

 In this subsection we describe the method developed for extracting the DiFFs using MC approach. Within this approach we  calculate various number densities by averaging over a large number of MC simulations of the quark hadronization process. These number densities can be expressed in terms of the corresponding fragmentation functions, allowing us to extract them from these number densities using the corresponding azimuthal modulations. In the case of production of two unpolarized hadrons by a longitudinally polarized quark, the relevant number density, according to Eqs.~(\ref{EQ_DELTA_UNP}) and (\ref{EQ_DELTA_LIN}), can be expressed as
\al
{
\label{EQ_F_VEC}
 F&(z, \xi , \kT, \RT; s_L) = D_1(z, \xi, \kT^2, \RT^2, \kT \cdot \RT) 
 \\ \non
 &+ s_L \frac{ (\RT \times  \kT)\cdot \vect{\hat{z}} }{M_1 M_2} G_1^\perp(z, \xi, \kT^2, \RT^2, \kT \cdot \RT) \, .
}
This can also be expressed in terms of the azimuthal angles $\fK$ and $\fR$ of the vectors $\kT$ and $\RT$
\al
{
\label{EQ_F_ANG}
F&(z, \xi , \kT, \RT; s_L) = D_1(z, \xi, \kT^2, \RT^2, \cos(\fRK)) 
 \\ \non
 &
 - s_L \frac{ R_T k_T \sin(\fRK) }{M_1 M_2} G_1^\perp(z, \xi, \kT^2, \RT^2, \cos(\fRK)).
}

The unpolarized and helicity dependent DiFFs of Eqs.~(\ref{EQ_D1_MH},\ref{EQ_G1_MH}) then can be extracted from the number density 
\al
{
D_1(z, M_h^2)=& \int d \xi  \int d \fR  \int d^2 \kT 
\\ \non 
 &\times \ F(z, \xi , \kT, \RT; s_L),
}
\al
{
G_1^\perp(z, M_h^2) = -&\frac{M_1 M_2}{s_L}  \int d \xi  \int d \fR  \int d^2 \kT 
 \\ \non 
 &\times \ \cot(\fRK)  F(z, \xi , \kT, \RT; s_L).
}

 In this work we concentrate on calculating $M_h^2$ integrated DiFFs, which are defined as
\al
{
 \label{EQ_D1_Z}
 D_1(z)
 = &  \int d \xi  \int d^2 \RT  \int d^2 \kT
 \\ \non 
 & \times \ D_1(z, \xi, \kT^2, \RT^2, \kT \cdot \RT),
 }
\al
{
 \label{EQ_G1_Z}
 G_1^\perp(z)
  =&  \int d \xi  \int d^2 \RT  \int d^2 \kT  
\\ \non
 & \times (\kT\cdot \RT) \ G_1^\perp(z, \xi, \kT^2, \RT^2, \kT \cdot \RT) \, .
}
The corresponding expressions in terms of the number density are
\al
{
\label{EQ_D1_Z_EXTR}
D_1(z)=& \int d \xi  \int d^2 \RT  \int d^2 \kT 
\\ \non 
 &\times \ F(z, \xi , \kT, \RT; s_L),
}
\al
{
\label{EQ_G1_Z_EXTR}
G_1^\perp(z) = -& \frac{M_1 M_2}{s_L} \int d \xi  \int d^2 \RT  \int d^2 \kT 
 \\ \non 
 &\times \ \cot(\fRK)   F(z, \xi , \kT, \RT; s_L).
}

Finally, throughout this work we will use the definition 
\al
{
 \tilde{G}_1^\perp(z) \equiv  \frac{1}{M_1 M_2} G_1^\perp(z) ,
}
to simplify the notation.

\subsection{The quark-jet model simulations}
\label{SUBSEC_NJL_JET}

\begin{figure}[!b]
\centering 
\includegraphics[width=0.8\columnwidth]{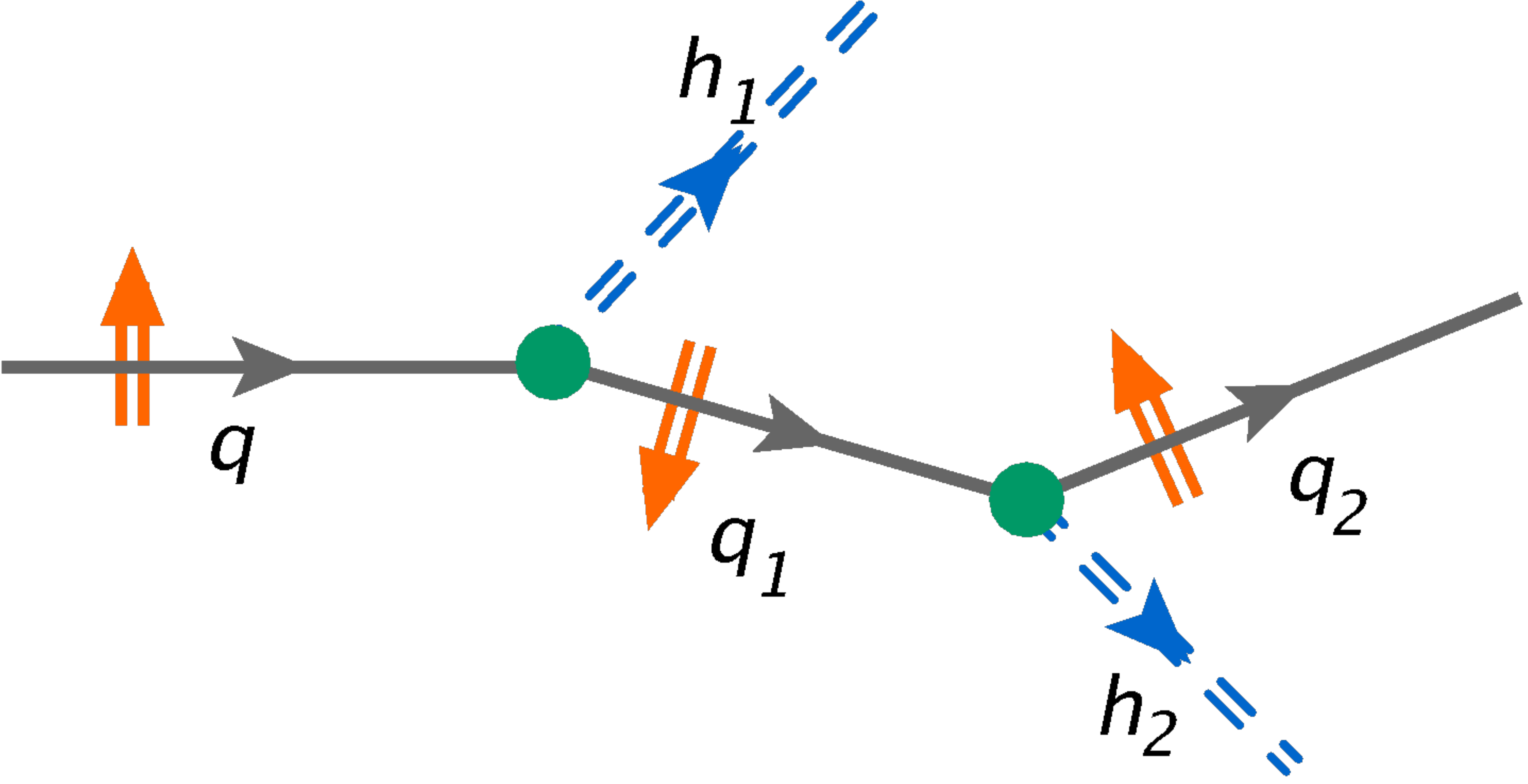}
\GapCapt
\caption{The extended quark-jet framework.}
\label{PLOT_QUARK_JET}
\end{figure}

In this subsection we describe the quark-jet MC simulation framework, used in the current study of the DiFFs, which has been developed over recent years~\cite{Matevosyan:2011ey,Matevosyan:2011vj, Matevosyan:2013aka, Matevosyan:2013nla, Matevosyan:2013eia, Matevosyan:2016fwi}.  We use the most recent evolution of the quark-jet framework, as detailed in Refs.~\cite{Bentz:2016rav, Matevosyan:2016fwi} to model the hadronization of a quark with a longitudinal polarization. The schematic depiction of the framework  in Fig.~\ref{PLOT_QUARK_JET} displays the sequential emission of hadrons $h_1$, $h_2$, etc., where the polarization of the remnant quark after each emission is determined by the corresponding spin density matrix. We choose a pre-determined number of hadron emissions, $N_L$, for terminating each hadronization chain and use MC to calculate the number densities for different hadron pairs. The input quark-to-quark SFs are calculated using the Nambu--Jona-Lasinio (NJL) effective quark interaction model~\cite{Nambu:1961tp,Nambu:1961fr}. The details of the model calculations and parameters are described in detail in Ref.~\cite{Matevosyan:2016fwi}. In this work, we forgo the QCD evolution of the DiFFs~\cite{Ceccopieri:2007ip}  we computed using the low-energy NJL effective model input. The evolution would be necessary for accurate direct comparisons  with the experimental results obtained at various large energy scales. We performed such studies in our previous work for the unpolarized DiFFs calculated within the same model in Ref.~\cite{Matevosyan:2013aka}, showing that the QCD evolution shifts the shape of the DiFFs towards lower $z$ region. To mimic such effects for the qualitative comparisons made in this article, we use the ansatz with all the input SFs multiplied by a factor of $(1-z)^4$. It is worth noting, that the particular choice of the input SFs does not affect the qualitative features of the computed DiFFs. Rather, the dominant aspects are the transverse momentum and spin transfer mechanisms within the quark-jet framework. All the results shown in this work are obtained by setting $s_L=1$, and $\sT{}=0$.

\begin{figure}[!b]
\centering 
\includegraphics[width=\ImL]{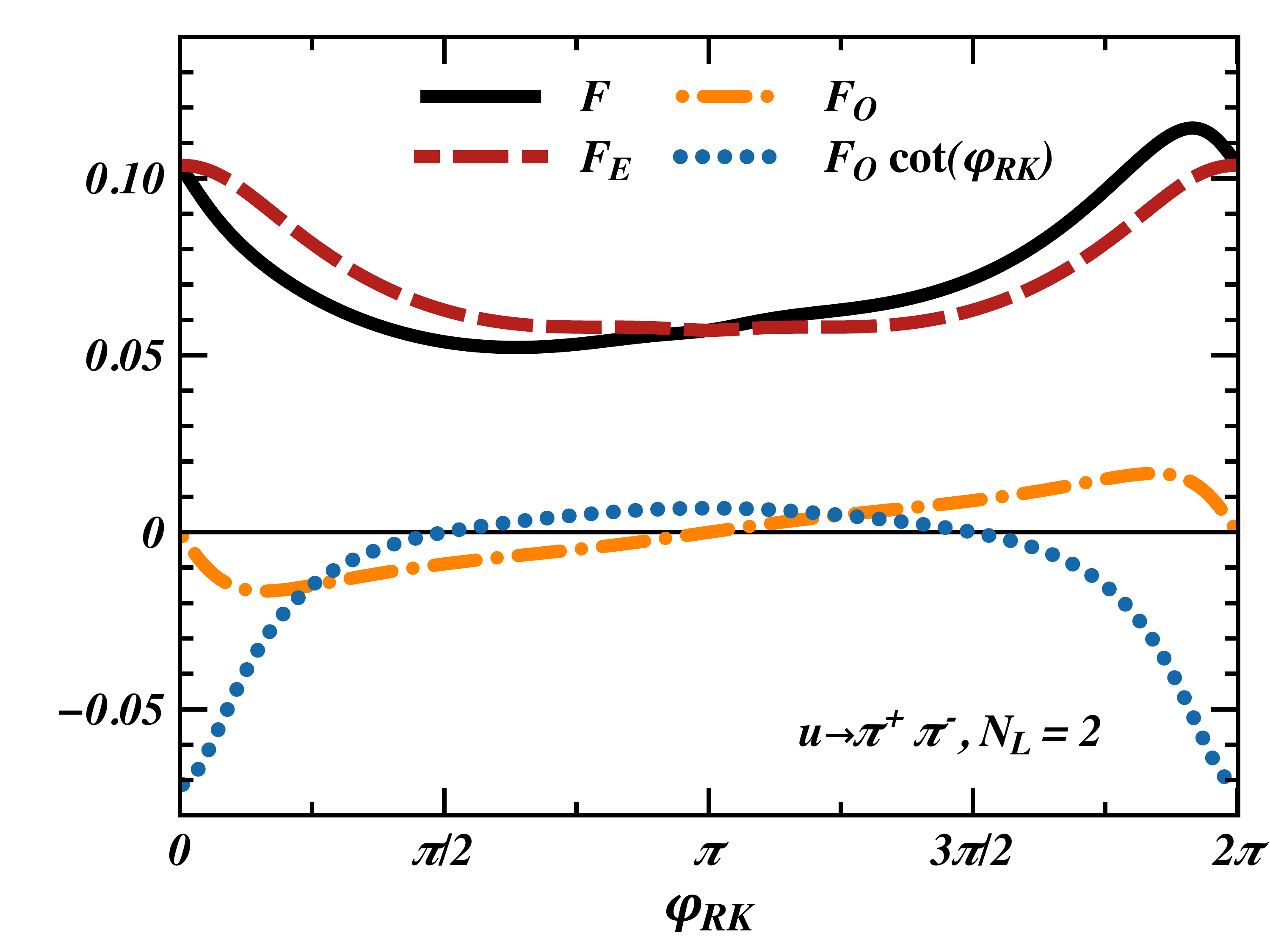}
\GapCapt
\caption{The MC results for the longitudinal spin correlations in $\pi^+\pi^-$ pairs for a hadronization of a $u$ quark with $N_L=2$. The horizontal axis represents the azimuthal angle $\fRK$, while the vertical axis represents the magnitudes of the $z$-integrated number densities and their modulations. The black solid curve depicts the total number density $F$, the red dashed and orange dash-dotted curves represent the even ($F_E$) and odd ($F_O$) parts of $F$ with respect to $\fRK$, and the blue dotted curve is the $F_O$ multiplied by $\cot(\fRK)$.}
\label{PLOT_PHIRT_MOD}
\end{figure}

Our first task is to test the method of extracting the DiFFs described in Sec.~\ref{SUBSEC_DIFF_EXTRACT}, considering $\pi^+\pi^-$ pairs produced with the smallest possible value of $N_L=2$. For this purpose,  in Fig.~\ref{PLOT_PHIRT_MOD} we show,using a black solid line, the $\fRK$ dependence of the function $F$ from~\Eq{EQ_F_ANG}, integrating over all the other variables. It obviously is not an even function of  $\fRK$, indicating a nonzero term associated with $G_1^\perp$ in \Eq{EQ_F_ANG}. The dashed and dash-dotted lines represent the even and the odd parts of this function, corresponding to the unpolarized and the helicity dependent DiFF terms, respectively. The blue dotted line is the odd part of $F$, multiplied by $\cot(\fRK)$. It is clear, that the integral of the red dashed and the blue dotted lines over $\fRK$ correspond to the $z$-integrated values of $D_1(z)$ and $\tilde{G}_1^\perp(z)$. Thus, we can ensure that the multiplication by the $\cot(\fRK)$ does not produce discontinuities for $\fRK = \{0, \pi\}$, yielding reliable estimates for $\tilde{G}_1^\perp$. We have to note though, that such extraction requires a large enough number of MC simulations to sufficiently suppress the statistical fluctuations. In calculating the $z$-dependence of $D_1$ and $\tilde{G}_1^\perp$, we use a similar procedure involving the $z$-unintegrated form of $F$. In this work we used $10^{12}$ MC events to calculate the polarized number densities, allowing us to reliably extract the DiFFs for $100$ discretization points of $z\in[0,1]$ and $200$ discretization points for $\fRK\in[0,2\pi)$.

\begin{figure}[!t]
\centering 
\subfigure[]
 {
\includegraphics[width=\ImL]{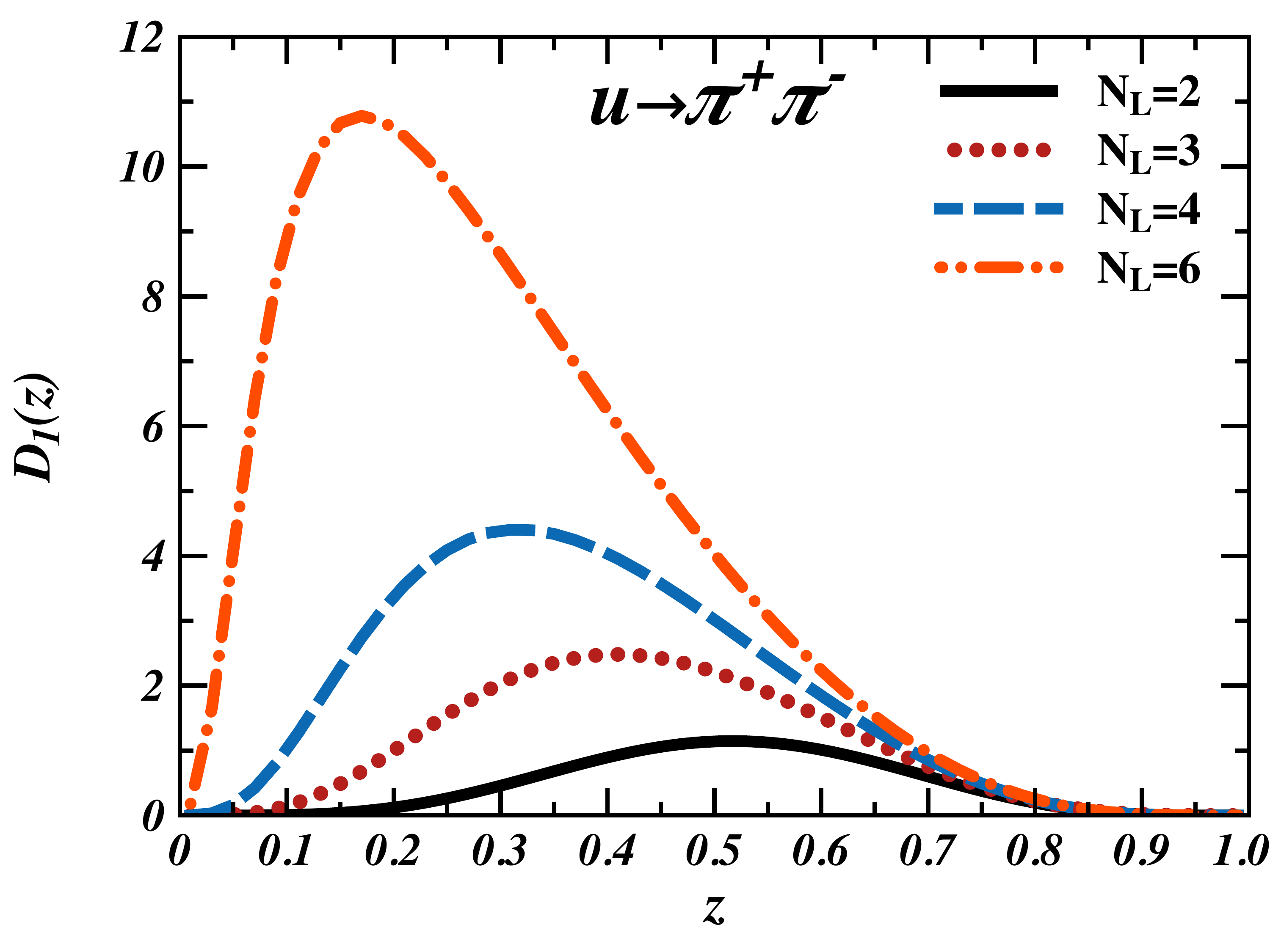}
}
\\ \GapSubf
\subfigure[]
 {
\includegraphics[width=\ImL]{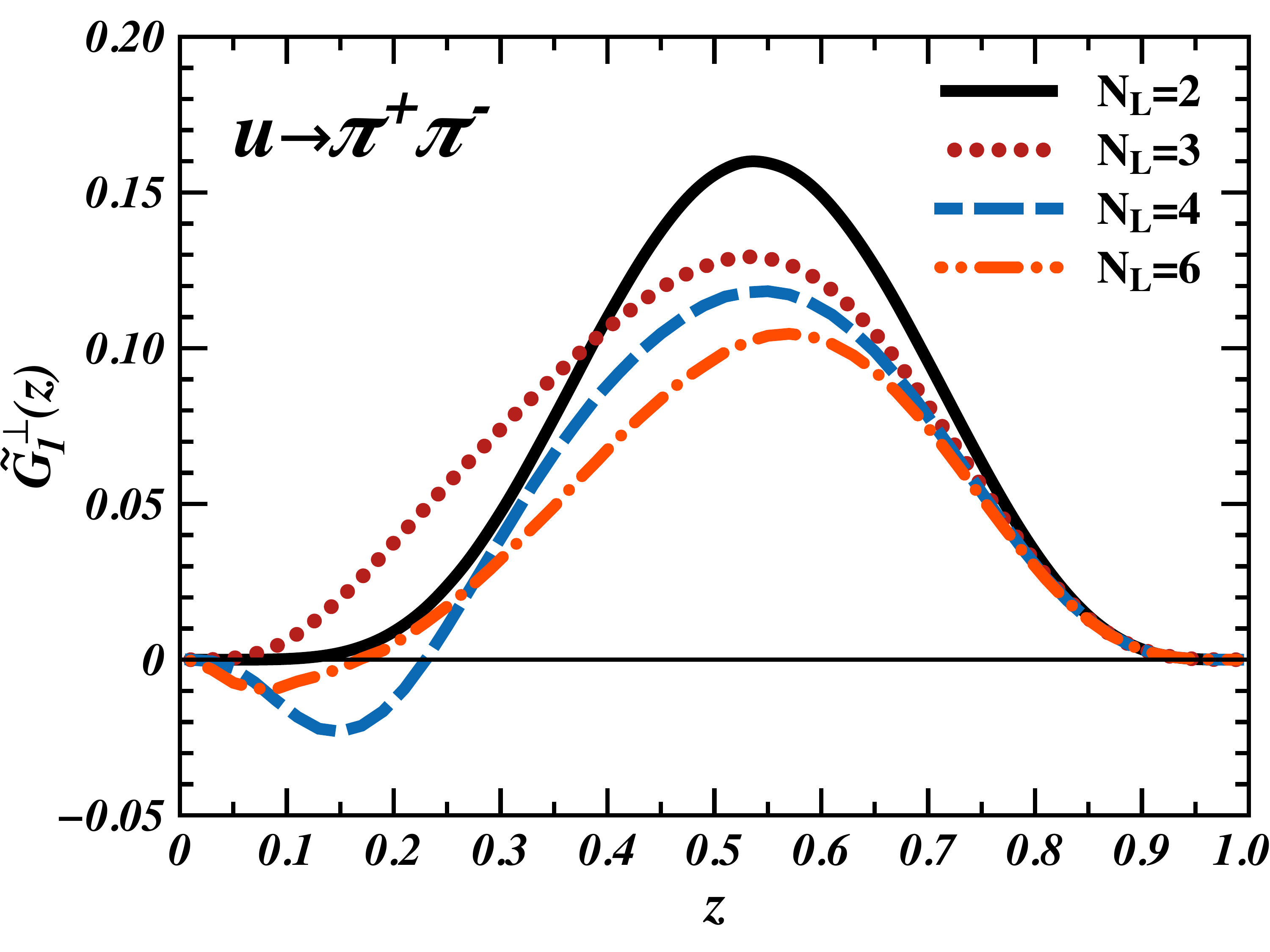}
}
\\ \GapSubf
\subfigure[]
 {
\includegraphics[width=\ImL]{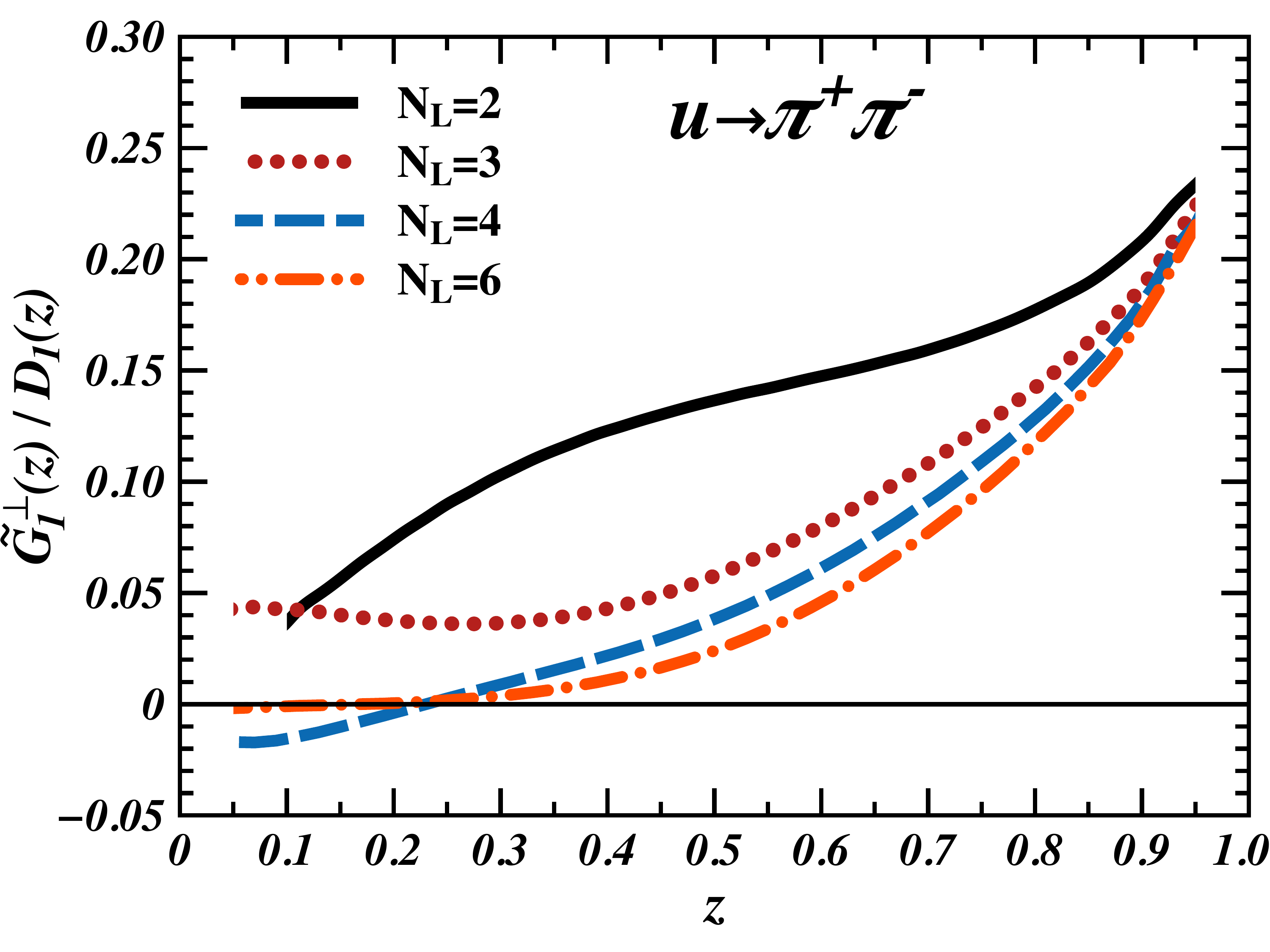}
}
\GapCapt
\caption{Comparison of MC results for $D_1(z)$ (a), $\tilde{G}_1^{\perp}(z)$ (b), and their ratios (c) for various values of $N_L$.}
\label{PLOT_D1_GP_PIPL_PIMI}
\end{figure}

\begin{figure}[!t]
\centering 
\subfigure[]
 {
\includegraphics[width=\ImL]{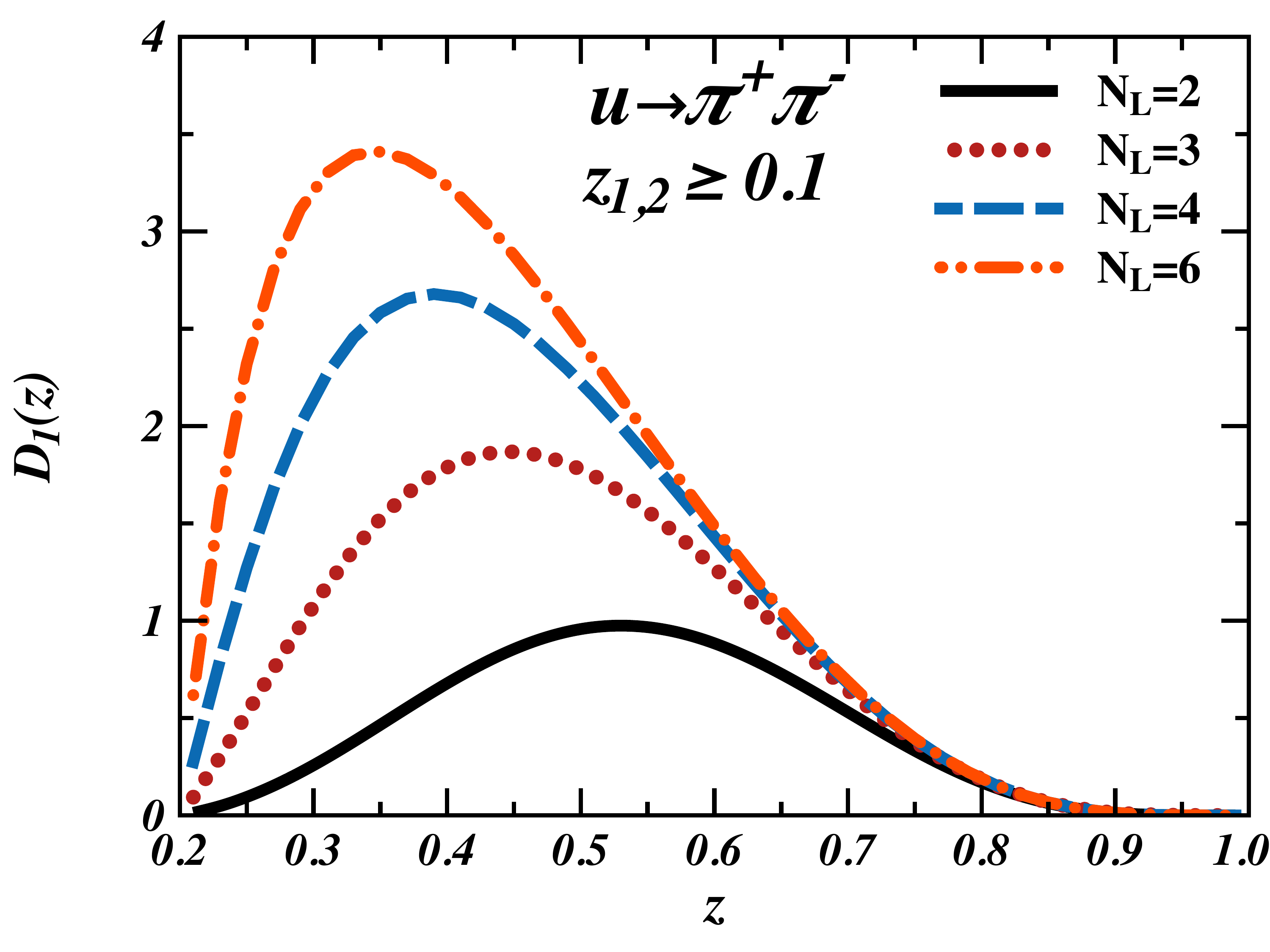}
}
\\ \GapSubf
\subfigure[]
 {
\includegraphics[width=\ImL]{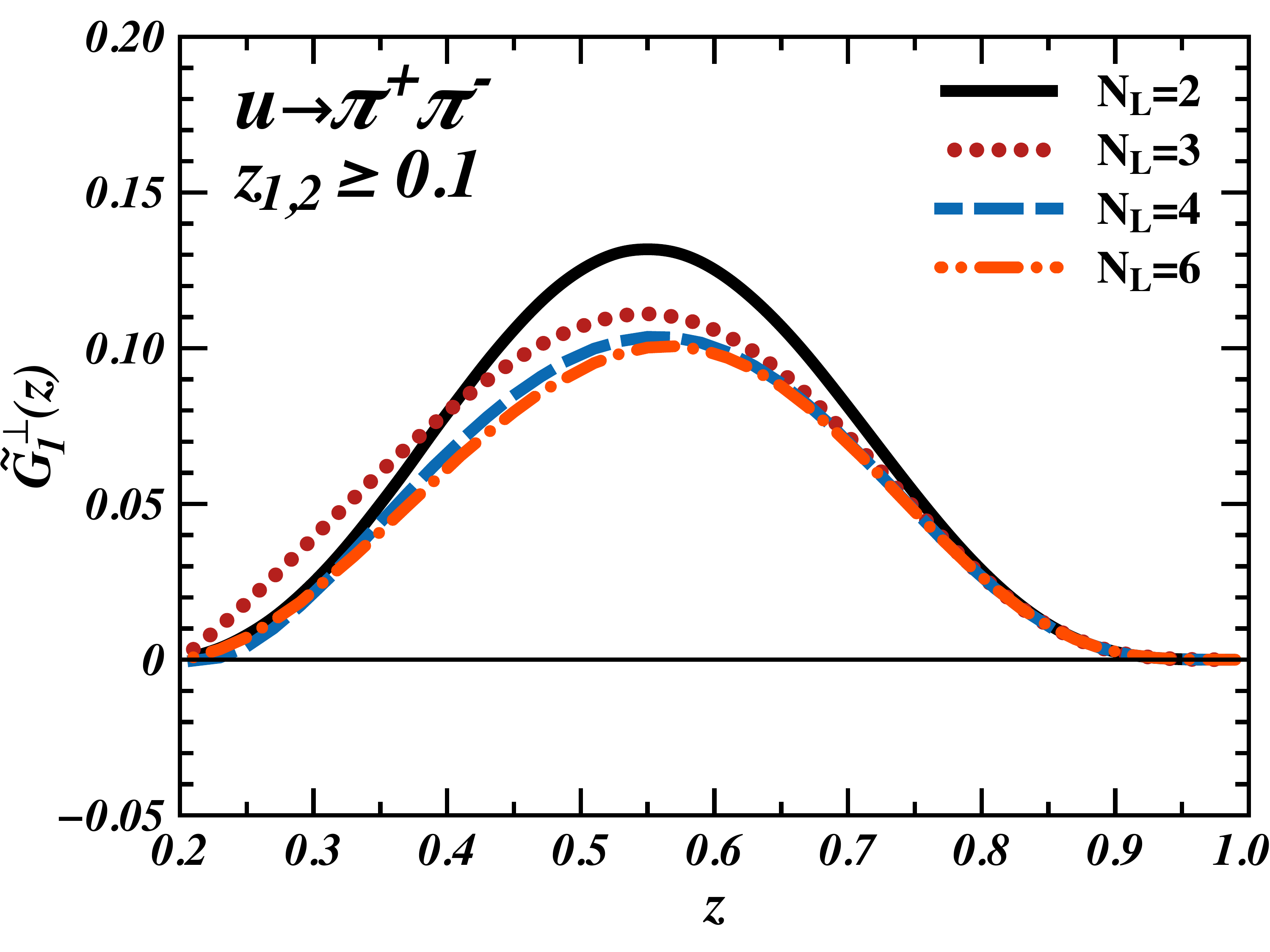}
}
\\ \GapSubf
\subfigure[]
 {
\includegraphics[width=\ImL]{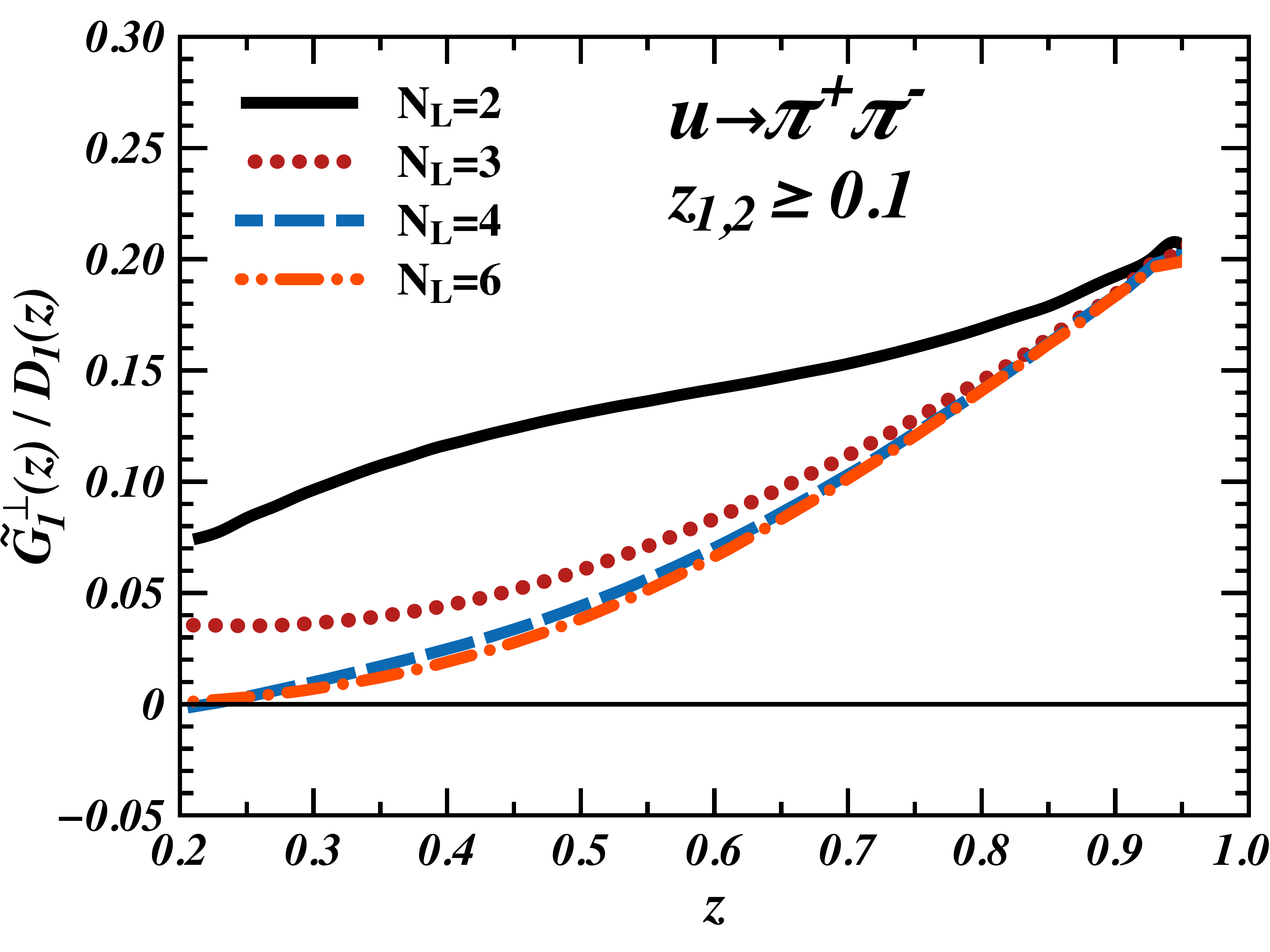}
}
\GapCapt
\caption{Comparison of MC results for $D_1(z)$ (a), $\tilde{G}_1^{\perp}(z)$ (b), and their ratios (c) for for various values of $N_L$. The  cuts $z_{1,2} \geq 0.1$ are imposed in these results.}
\label{PLOT_D1_GP_PIPL_PIMI_ZCUT}
\end{figure}

The results for the $z$-dependent extractions of $D_1$, $\tilde{G}_1^\perp$, and their ratios of the $\pi^+\pi^-$ pairs are depicted in Fig.~\ref{PLOT_D1_GP_PIPL_PIMI}. Here the black solid , the red dotted , the blue dashed and the orange dash-dotted lines depict the results for $N_L=2,3,4,$ and $6$, respectively. It is clear that the function $D_1(z)$ increases rapidly in the small-$z$ region with increasing number of produced hadrons $N_L$, due to the large combinatorial factors in selecting pairs from an expanding set of final particles. It is also remarkable to see a nonvanishing signal for $\tilde{G}_1^\perp(z)$, which gets smaller with an increasing number of produced hadrons because of the destructive interference of oppositely signed signals for the pairs produced at different ranks. Nevertheless, the results for the "analyzing power" of the $\tilde{G}_1^\perp(z)$ modulations of the number density $F$, depicted in Fig.~\ref{PLOT_D1_GP_PIPL_PIMI}(c), show that only in the large-$z$ region is there a significant signal that can be possibly measured.

The plots in  Fig.~\ref{PLOT_D1_GP_PIPL_PIMI_ZCUT} are analogous to those in Fig.~\ref{PLOT_D1_GP_PIPL_PIMI}, except that here an additional cut is placed upon the minimum values of $z$ for each member of the pair, $z_{1,2}\geq 0.1$. Such cuts are common in experimental settings, and for example in SIDIS are mainly aimed at separating the current and target fragmentation regions. Thus it is important to evaluate the impact on our results. The plots show the expected suppression of the unpolarized DiFF with respect to the previous case, as the hadrons with increasing rank on average carry decreasing values of light-cone momentum $z$, thus the high-ranked hadrons are often excluded from the pairs due to the cut criterion. On the other hand, the impact on the helicity dependent DiFF is less significant, since the bulk of the effect is generated by the first two produced hadrons. This results in a slight increase of the analyzing power in the mid-$z$ range, making it less peaked towards large $z$.

\begin{figure}[!b]
\centering 
\subfigure[]
 {
\includegraphics[width=\ImL]{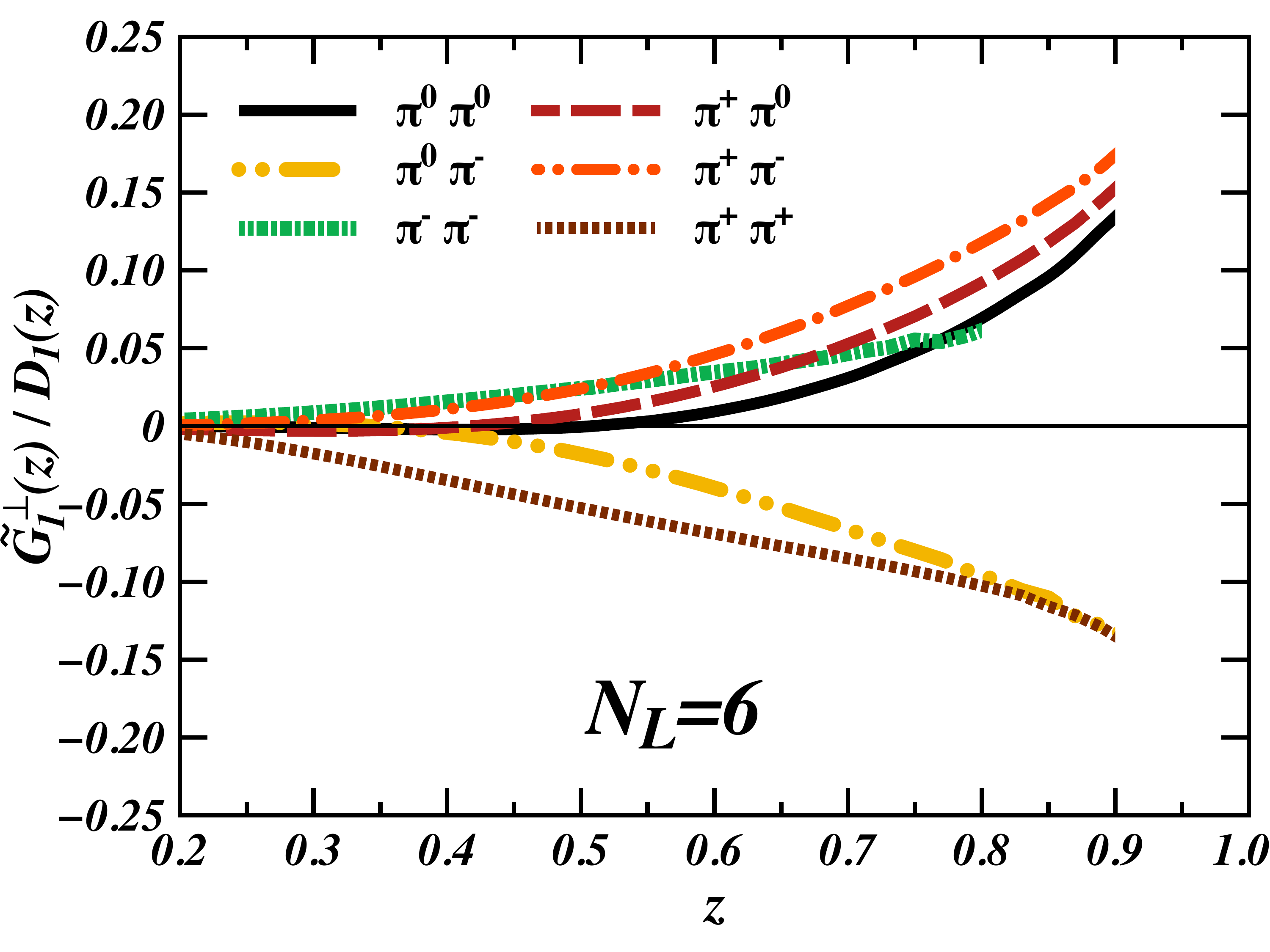}
}
\\ \GapSubf
\subfigure[]
 {
\includegraphics[width=\ImL]{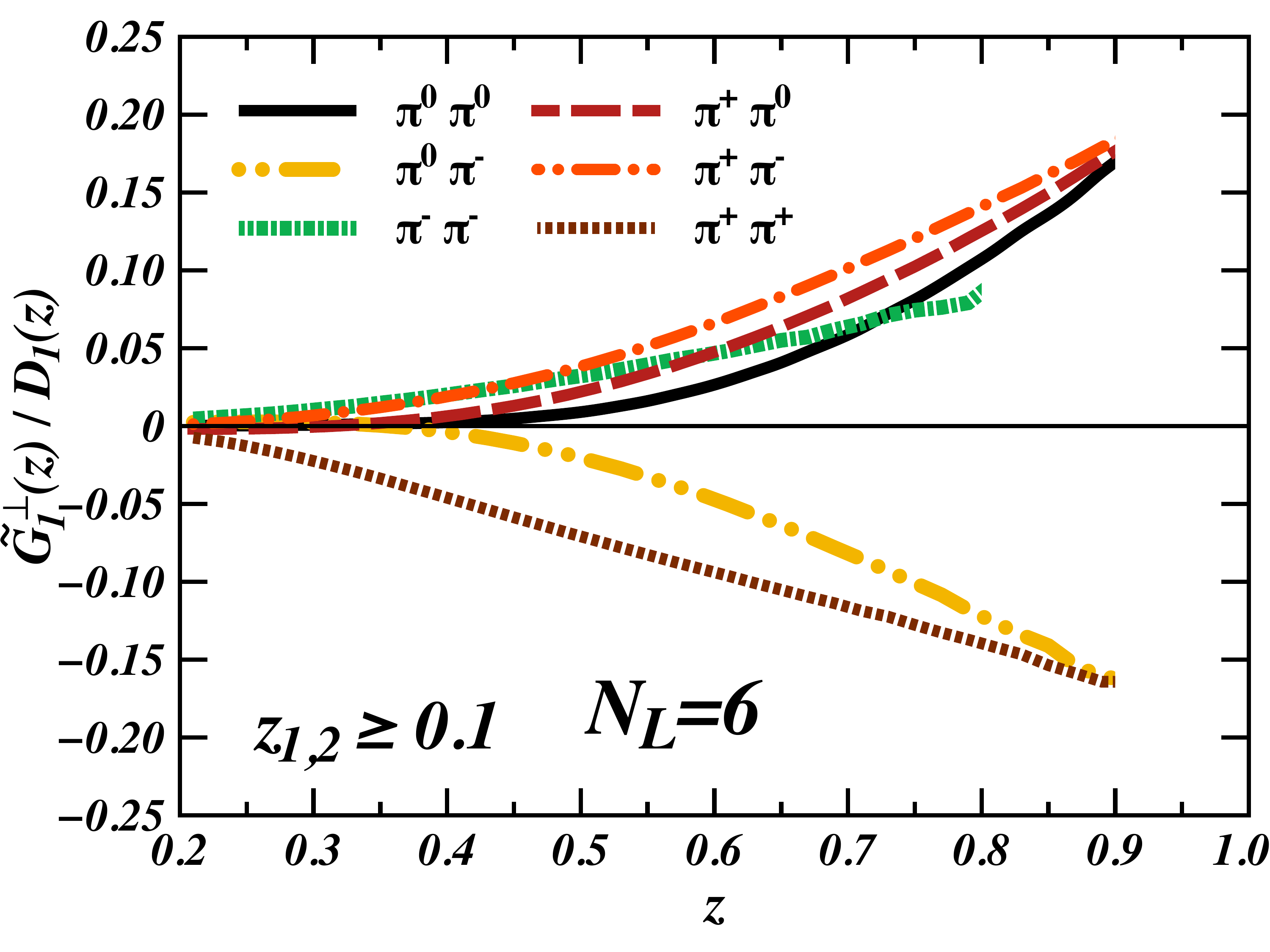}
}
\GapCapt
\caption{The ratio $\tilde{G}_1^{\perp}(z)/D_1(z)$ for different pion pairs (a),  and the cuts of $z_{1,2} \geq 0.1$ (b),  for simulations with $N_L=6$.}
\label{PLOT_GP_RAT_PAIRS}
\end{figure}

 Next, we plot the results for all possible types of pion pairs in Fig.~\ref{PLOT_GP_RAT_PAIRS} for $N_L=6$, both without and with the $z$ cut. It is interesting to see, that $\pi^+\pi^+$ pairs have an opposite sign and similar magnitude to the $\pi^+\pi^-$ case. Here, in the same-signed pairs we assign the hadron with the larger $z$ as the first member of the pair. Without such a choice for $z$-ordering, all the results for the same-signed pairs vanish, as expected from the symmetry considerations.
 
\begin{figure}[!t]
\centering 
\includegraphics[width=\ImL]{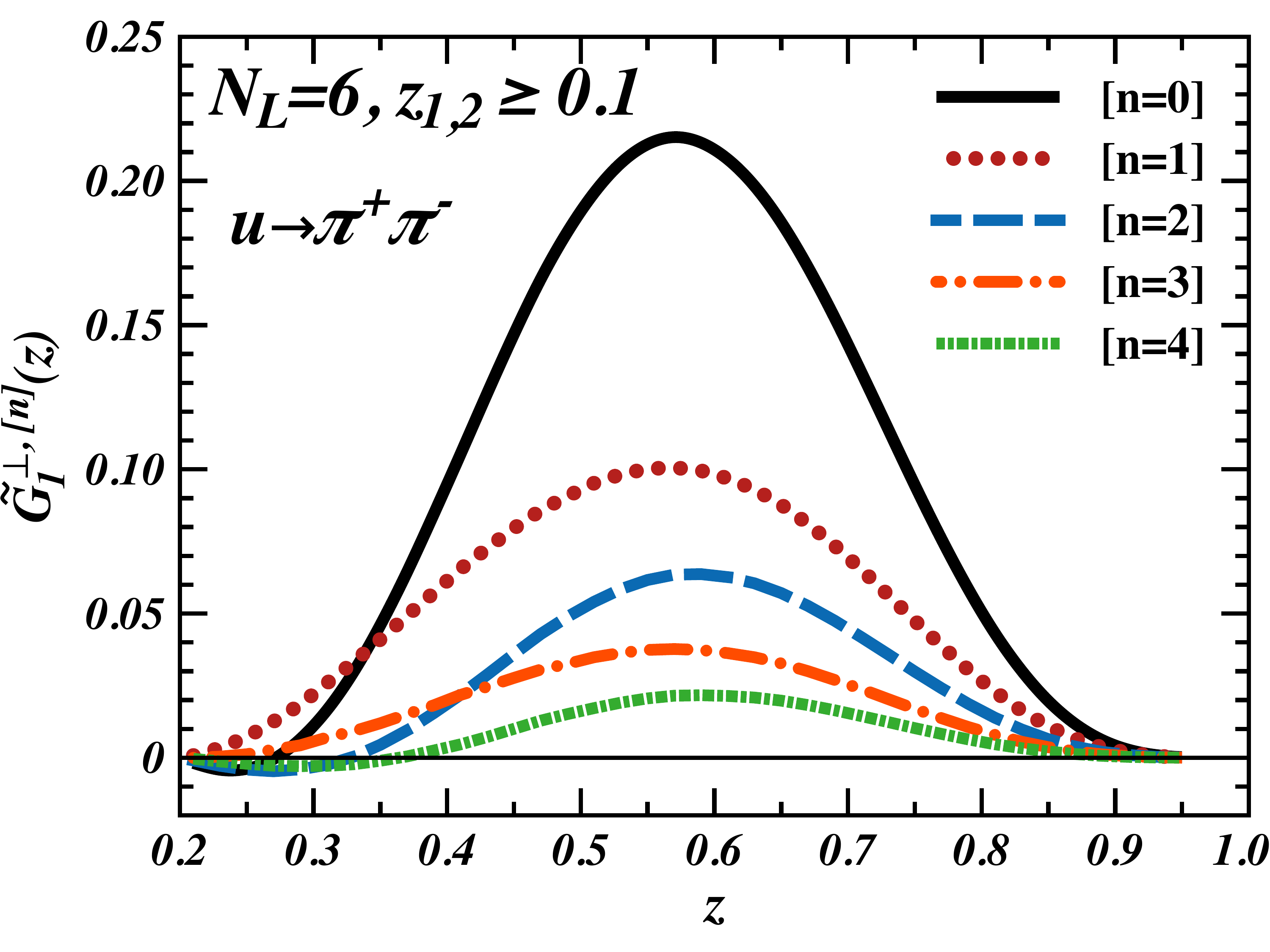}
\GapCapt
\caption{The first five Fourier cosine moments of $\tilde{G}_1^{\perp}(z)$ for $\pi^+\pi^-$ pairs and $N_L=6$, imposing  the cuts of $z_{1,2} \geq 0.1$.}
\label{PLOT_GP_MOM_PIPL_PIMI_ZCUT}
\end{figure}

The SIDIS cross  section for two hadron production contains all the cosine moments of $G_1^{\perp}$. In  Fig.~\ref{PLOT_GP_MOM_PIPL_PIMI_ZCUT} we we present the first five moments for $\pi^+\pi^-$ and $N_L=6$, extracted from the polarized number density in a analogous manner to the first moment. Here only the results with cuts  $z_{1,2} \geq 0.1$ are presented.
 
\section{MC Validation}
\label{SEC_VALIDATION}

In this section we validate our MC simulations in two ways. First, we use combinatorial arguments to calculate the total number of all $\pi^+\pi^-$ pairs  for a given $N_L$ and compare these calculations to the results from MC simulations. Second, we derive explicit integral relations for the $z$ dependence of the DiFFs for the case  $N_L=2$, and compare them with the MC results.

\subsection{Unpolarized DiFFs - Total number of dihadron pairs}

It is useful to calculate the number of the dihadron pairs of a given type ($\pi^+ \pi^-$, $\pi^0\pi^0$,  etc) for a fixed number of produced hadrons, where we integrate over all variables. Here we will limit ourselves to consider only pions produced by the $u,d$ quarks. Let us look at the initial $u$ quark producing $N_L$ pions, and at each step the total probability of producing a charged pion has an isospin factor of $1$, while the neutral pion has a isospin factor of $\frac{1}{2}$.  In the quark-jet framework, the total probability of producing a hadron at each step is $1$, then the probability of producing a charged pion is $\frac{2}{3}$, while for a neutral is $\frac{1}{3}$.

 It is important to note that, because of flavor conservation, the first produced charged pion should be a $\pi^+$, and each subsequent charged pion should have an alternate charge. Neutral pions can be produced at any stage without such limitations. First, we count the number of different cases when producing $n_0$ neutral pions, where $0\leq n_0 \leq N_L$.  Let us start with one of the possible scenarios for $n_0$ produced hadrons
\al
{
 (\overbrace{\pi^+, \pi^-, \pi^+, ...,}^{N_L- n_0} \overbrace{\pi^0, \pi^0, \pi^0}^{n_0}).
}
The number of all possible permutations of this set is  $N_L!$. Thus number will be over-counting the cases with all permutations of just  $\pi^0$s, which is $n_0!$. Also, for a given $n_0$, there is only a single ordering of the charged pion possible, as a permutation of any two same charged hadrons would yield an identical set, while  a permutation of opposite signed hadrons would yield an invalid set that violates flavor conservation. Thus, we should also divide the total number of sets by all possible permutations of the charged pions, that is $(N_L - n_0)!$. Then, the number of such combinations is 
\al
{
  N_{n_0} = \frac{N_L!}{n_0 ! (N_L -n_0)!} \equiv C^{n_0}_{N_L}.
}
and the probability of producing any such combination is 
\al
{
  P(n_0) = \Big(\frac{2}{3} \Big)^{N_L - n_0} \Big(\frac{1}{3} \Big)^{ n_0}.
}
Just a quick verification of our formula can be obtained by calculating the total probability of producing $N_L$ hadrons (all possible combinations) in $N_L$ steps, given by
\al
{
 P &= \sum_{n_0 =0}^{n_0 = N_L} N_{n_0}  P(n_0) = 1,
}
as expected.

The number of various $\pi \pi$ pairs in each combination is
\al
{
&
 N^ {(\pi^+ \pi^-)}(n_0) = U\Big( \frac{N_L - n_0}{2} \Big) \ D\Big( \frac{N_L - n_0}{2} \Big),
\\
&
N^ {(\pi^0 \pi^+)}(n_0) = n_0\ U\Big( \frac{N_L - n_0}{2} \Big) ,
\\
&
N^ {(\pi^- \pi^0)}(n_0) =  D\Big( \frac{N_L - n_0}{2} \Big) \ n_0,
\\
&
 N^ {(\pi^+ \pi^+)}(n_0) =  C_{U\Big( \frac{N_L - n_0}{2} \Big)}^2,
\\
&
 N^ {(\pi^- \pi^-)}(n_0) = C_{D\Big( \frac{N_L - n_0}{2} \Big)}^2,
 \\
&
 N^ {(\pi^0 \pi^0)}(n_0) = C_{n_0}^2,
}
where the $U(n), D(n)$ functions round up and down to the nearest integer.

Finally, the mean number of producing a $\pi^+\pi^-$ pair after $N_L$ emissions is
\al
{
\label{EQ_N_PIPI}
 &\mathcal{N}^ {(\pi^+ \pi^-)}(N_L) = \sum_{n_0 =0}^{n_0 = N_L} N_{n_0}  P(n_0)   N^ {(\pi^+ \pi^-)}(n_0)
 \\ \non
 &=     \sum_{n_0 =0}^{n_0 = N_L} C^{n_0}_{N_L} \Big(\frac{2}{3} \Big)^{N_L - n_0} \Big(\frac{1}{3} \Big)^{ n_0} U\Big( \frac{N_L - n_0}{2} \Big) \ D\Big( \frac{N_L - n_0}{2} \Big).
}

 It is also clear that this number is simply the integral over $z$ of the unpolarized DiFF extracted from MC simulations with $N_L$ produced hadrons.
\al
{
\label{EQ_D1_INT}
   \mathcal{N}^ {(\pi^+ \pi^-)}_{MC}(N_L) = \int_0^1 dz \ D_{1, [N_L]}^{u\to\pi^+\pi^-}(z)
}

The results of the calculations both using \Eq{EQ_N_PIPI} and \Eq{EQ_D1_INT} for a range of values of $N_L$ are presented in Table~\ref{TABLE_N_PIPI}. We see very good agreement between the two methods, given the discretization errors of the MC simulations. The last row in Table~\ref{TABLE_N_PIPI} represents the results of MC simulations with cuts on minimum value of $z$ for each hadron in the pair, $z\geq z_{min}=0.1$. 
\begin{table}[tb]
\centering
\begin{center}
\begin{tabular}
{ | C{0.5cm} |  C{1.5cm} | C{1.5cm}  | C{1.5cm}  | C{1.5cm}   m{-30pt} |}
\hline
 $N_L$ & $\mathcal{N}^ {(\pi^+ \pi^-)}$ & $\mathcal{N}^ {(\pi^+ \pi^-)}_{N} $ & $\mathcal{N}^ {(\pi^+ \pi^-)}_{MC}$ & $\mathcal{N}^ {(\pi^+ \pi^-)}_{MC, z_{min}}$ & \\  [2.5ex] \hline \hline
 2 & $\dfrac{4}{9}$			& 0.44444		& 0.4444		  	& 0.350175	& 	\\ [3.5ex] \hline
 3 & $\dfrac{28}{27}$ 		& 1.03704		& 1.03694 		& 0.683999	&	\\ [3.5ex]\hline
 4 & $\dfrac{152}{81}$ 		& 1.87654		& 1.87641 		& 0.959588	&	\\ [3.5ex]\hline
 5 & $\dfrac{712}{243}$	 	& 2.93004		& 2.92992 		& 1.11531		&	\\ [3.5ex]\hline
 6 & $\dfrac{3068}{729}$ 		& 4.2085 		& 4.20882 		& 1.18162 	&	\\ [3.5ex]\hline
 7 & $\dfrac{12484}{2187}$ 	& 5.70828		& 5.70867 		& 1.20282	 	&	\\ [3.5ex]\hline
 8 & $\dfrac{48752}{6561}$ 	& 7.43057		& 7.43047 		& 1.20809		&	\\ [3.5ex]\hline
\end{tabular}
\caption{The number of $\pi^+\pi^-$ pairs $\mathcal{N}^ {(\pi^+ \pi^-)}$ for a given number of produced hadrons $N_L$. The numbers in the third row are the approximate numerical values obtained via the \Eq{EQ_N_PIPI}, while those in the fourth row are the results of the numerical simulations and \Eq{EQ_D1_INT}.  The last row shows the results form the same MC simulations with cuts on minimum value of $z$ for each hadron in the pair: $z_{1,2}\geq z_{min}=0.1$.}
\label{TABLE_N_PIPI}
\end{center}
\end{table}

\subsection{Two-step process and validation}
\label{SUBSEC_VALID_TWO}

Here we aim to validate our MC results for both unpolarized and helicity dependent DiFFs by deriving explicit integral relations when an initial $u$ quark produces only a single $\pi^+\pi^-$ pair. We used a similar approach in sections IIC and IVA  of~\cite{Matevosyan:2016fwi} to derive similar expressions for the unpolarized and unfavored Collins function for a $u\to\pi^-$ fragmentation for the case of  two-hadron emission. We briefly review the kinematics setup here, and a more detailed description can be found in Ref.~\cite{Matevosyan:2016fwi}. 

 In the quark-jet framework, the fragmenting quark $q$ (the initial $u$ quark), emits a hadron $h_1$ (a $\pi^+$ in our calculations), leaving a remnant quark $q_1$ (a $d$ quark) carrying light-cone momentum fraction $\eta_1$ and transverse momentum component $\pe{1}$, where the transverse direction is defined with respect to the three-momentum vector of $q$. The initial and remnant quark's spin 3-vectors are denoted as $\vect{s}_q = (\vect{0},s_L )$ and $\vect{s}_{q_1} = (\vect{s}_{T_1}, s_{L_1})$. In the second hadronization step, the quark $q_1$ emits a hadron $h_2$ (a $\pi^-$) with light-cone momentum fraction $\eta_2$ and transverse momentum $\pe{2}$ with respect to the 3-momentum direction  of $q_1$. We can then easily calculate the momentum of $h_2$ in the initial frame using the Lorentz transformation in \Eq{EQ_LORENTZ}.The number density for the process of $u\to \pi^+\pi^-$ is simply the product of the corresponding number densities for each of these two steps
\al
{
  \label{EQ_Q_to_2H}
F^{(2)}_{q\to h_1 h_2}(&\eta_1, \pe{1}, \eta_2, \pe{2}; \sq{q} )
\\ \non
=& \ \hat{f}^{q\to q_1 }(\eta_1, \pe{1};  \sq{q} )  \cdot \hat{f}^{q_1\to h_2}(\eta_2, \pe{2}; \sq{q_1} ) \, ,
}
where the elementary probability densities can be expressed in terms of the elementary TMD SFs
\al
{
\label{EQ_FHAT_Q_Q1}
 \hat{f}^{q\to q_1 }(z, \pe{}; & \vect{s} )
 \\ \non 
 = \hat{D}^{(q\to q_1)}&(z, \psq{})  + \frac{(\pe{}\times \sT) \cdot \hat{\vect{z}} }{z M_{q_1}}  \ \hat{H}^{\perp({q\to q_1})}(z,\psq{}),
}
\al
{ 
\label{EQ_FHAT_Q_H}
  \hat{f}^{q \to h}(z, \pe{};& \vect{s} )  
\\ \non
= \hat{D}^{(q\to h)}&(z, \psq{})  + \frac{(\pe{}\times \sT) \cdot \hat{\vect{z}}}{z m_h}\ \hat{H}^{\perp({q\to h})}(z,\psq{}),
}
The remnant quark's polarization is determined using the quark spin density matrix formalism, which for an initial longitudinally polarized quark reads
\al
{
 \vect{s}_{T_1}  =& \frac{1}{ \hat{f}^{q\to q_1}(\eta_1, \pe{1};  \sq{q}) } 
 \\ \non
 & \times \Bigg( \frac{ \pe{1}' }{\eta_1 M_{q_1}}  \hat{D}_T^\perp(\eta_1,\psqn{1})  - \frac{\pe{1} }{\eta_1 M_{q_1}} s_L  \hat{G}_T(\eta_1,\psqn{1}) \Bigg),
  \\
 s_{L_1} = & \frac{1 }{ \hat{f}^{q\to q_1}(\eta_1, \pe{1};  \sq{q}) } s_L \ \hat{G}_L(\eta_1,\psqn{1}),
}
and $\pe{1}' \equiv (-p_{1y},p_{1x})$. Here $\hat{D}$, $\hat{D}_T^\perp$,$\hat{G}_L$,  $\hat{G}_T$, and $\hat{H}^{\perp}$ are the TMD elementary SFs. A quark model calculation of all the eight quark-to-quark and the two quark-to-hadron TMD SFs has been done in Ref.~\cite{Matevosyan:2016fwi} using the spectator approach. Note, that only $\vect{s}_{T_1} $ contributes to $F^{(2)}_{q\to h_1 h_2}$, as can be seen from \Eq{EQ_FHAT_Q_H}. 

The momenta of $h_1$ and $h_2$ are obtained using the momentum conservations and the Lorentz transformations similar to that in \Eq{EQ_LORENTZ} to calculate the transverse momenta of $h_2$ in the initial quark's system
\al
{
 & z_1 = 1-\eta_1,
&\Pe{1}= - \pe{1},
\\
&z_2 = \eta_1 \eta_2,
&\Pe{2} = \pe{2} + \eta_1\pe{1}.
}

 The final results for the $z$-dependence of the unpolarized and helicity dependent DiFFs are
\al
{
\label{EQ_D1_NL2}
 D_1^{(2)}(z) 
 = \int_{0}^{1} d \eta_1 \int_{0}^{1} & d \eta_2
\  \delta \Big(z - 1 +\eta_1 (1-\eta_2)  \Big)
 \\ \non
& \times  \ \hat{D}^{q\to q_1}(\eta_1)  \ \hat{D}^{q_1\to h_2} ( \eta_2 ),
}
and
\al
{
\label{EQ_GP_NL2}
\non
&\tilde{G}_1^{\perp (2)}(z)
=   -\pi^2 \int_{0}^{1} d \eta_1 \int_{0}^{1} { d \eta_2}  \  \delta \Big(z - 1 +\eta_1 (1-\eta_2)  \Big)
\\ 
& \ \times
\ \int d p_{1\perp}^2 \int d p_{2\perp}^2
  \frac{ \eta_2 (1- \eta_2)\psqn{1} -  (1-\eta_1)\psqn{2} }{z}
  \\ \non
& \ \times 
 \   \frac{1 }{\eta_1M_{q_1} }  \hat{G}_T^{q\to q_1}(\eta_1,\psqn{1})
    \  \frac{1 }{\eta_2 M_2} \ \hat{H}^{\perp (q_1\to h_2)}(\eta_{2},\psqn{2}).
}
It is important to note, that the above results are obtained only using the quark-jet formalism and the spin transfer mechanism, and do not depend on the underlying quark models used to calculate the particular forms of the SFs.

\begin{figure}[t]
\centering 
\subfigure[]
 {
\includegraphics[width=\ImL]{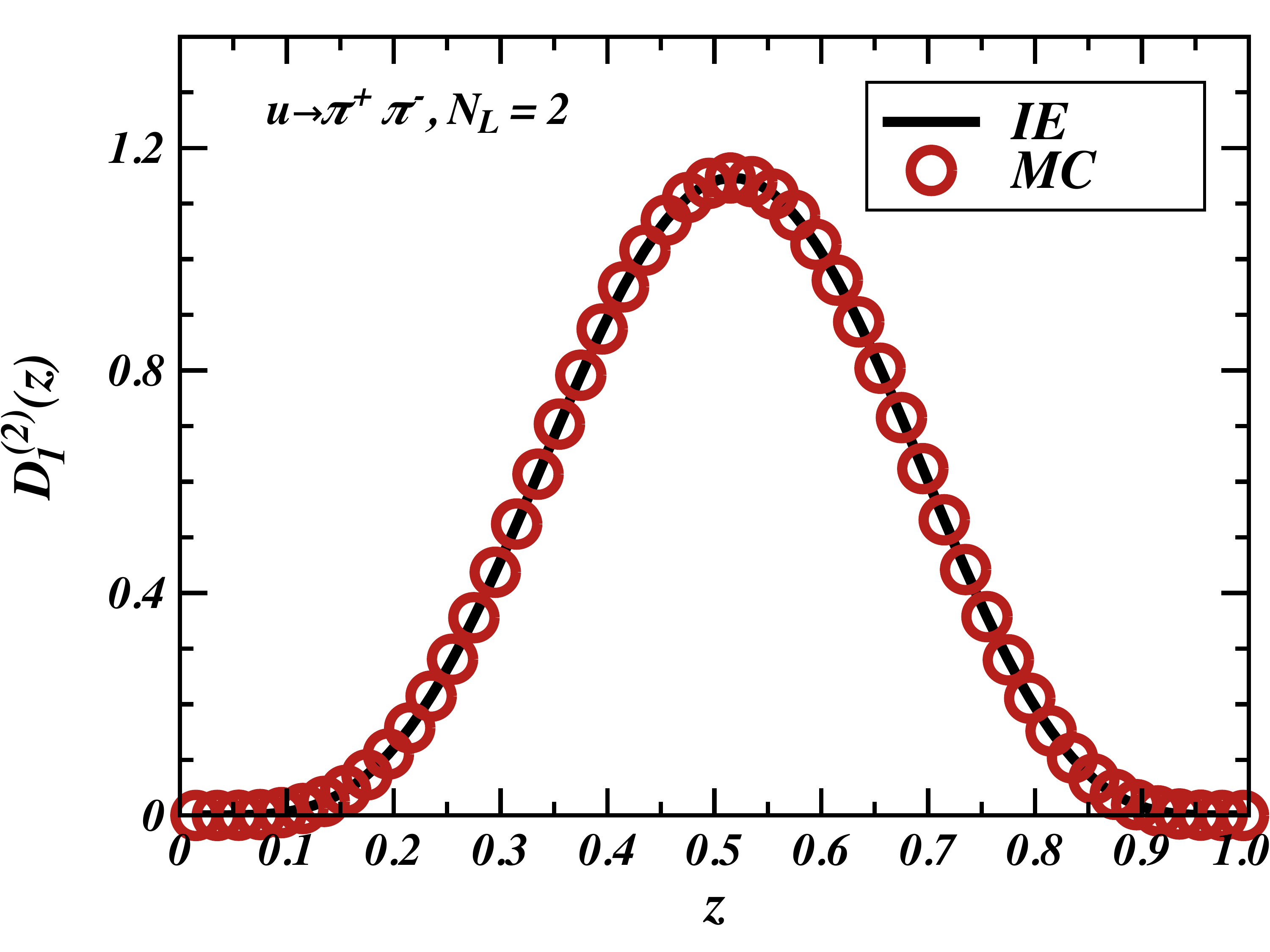}
}
\\ \GapSubf
\subfigure[]
 {
\includegraphics[width=\ImL]{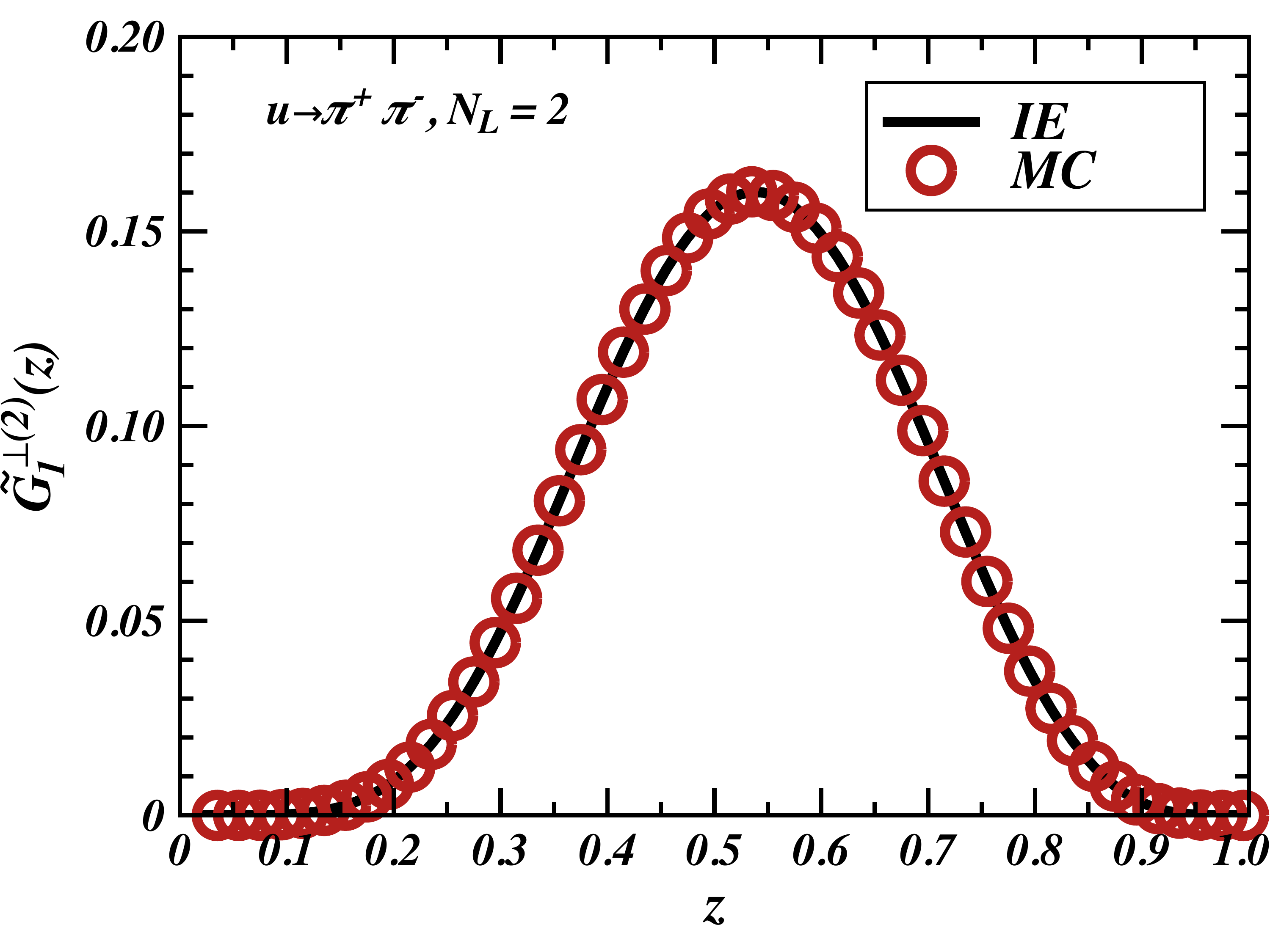}
}
\GapCapt
\caption{Comparison of results of  $D_1(z)$ (a), and  $\tilde{G}_1^{\perp}(z)$ (b) of integral expressions (IE) and MC results for $\pi^+\pi^-$ pair produced for $N_L=2$ in the quark-jet picture.}
\label{PLOT_D1_GP_NL2}
\end{figure}

The plots in Fig.~\ref{PLOT_D1_GP_NL2} depict the results for the unpolarized DiFF (a) and  the helicity dependent DiFF (b), calculated both using the MC method (plotted with red points) and the explicit integral relations in Eqs.~(\ref{EQ_D1_NL2},\ref{EQ_GP_NL2}) (plotted with black lines) for $N_L=2$. We observe excellent agreement between the two methods, both for the unpolarized and the helicity dependent DiFFs, validating our calculations.  

 It is also worth examining the structure of \Eq{EQ_GP_NL2}, which elucidates the microscopic mechanism for generating the helicity dependent DiFFs in the quark-jet framework. Here the "worm-gear" type elementary splitting function $G_T$ for the first quark-to-quark process, describing the correlation of the transverse polarization of the remnant quark with the longitudinal polarization of the fragmenting quark, is convoluted with the elementary Collins function $H^\perp$ for the second hadron emission. The latter describes the correlation of the emitted hadron's transverse momentum with the quark's transverse polarization. Thus, even in the case of longitudinal polarization, the Collins effect in the single hadron emission process, together with the momentum recoil mechanism of the quark-jet framework, is responsible for generating two-hadron correlations with the initial longitudinal polarization. This is a fascinating result, as naively the Collins effect is associated with the correlations involving transverse polarization. Finally, we have to note that it is possible for a helicity dependent two hadron correlation to be generated by the  so-called interference mechanism of the hadron pair produced in different channels (let us say as decay products of resonances emitted by the quark), as detailed in Ref.~\cite{Bianconi:1999cd}. Nevertheless, such calculations are beyond the scope of this work.

\section{Conclusions}
\label{SEC_CONCLUSIONS}

Spin-dependent correlations in two-hadron fragmentation functions provide a wealth of information about the hadronization process. Moreover, they provide an additional method for exploring the momentum and spin structure of the nucleon via two-hadron SIDIS measurements. Nonetheless, these DiFFs are still not very well known. This is especially true for the helicity dependent DiFF $G_1^\perp$, as the most recent efforts to measure the asymmetries involving this function in back-to-back two hadron production in $e^+e^-$ annihilation produced no significant signal.
 
In this work we have described the helicity dependent two-hadron correlations in hadronization of a longitudinally polarized quark using the quark-jet hadronization framework. We derived the method for extracting the $G_1^\perp$ function from the number density of hadron pairs produced by a polarized quark in Sec.~\ref{SUBSEC_DIFF_EXTRACT}. Then, we used the NJL model input quark TMD splitting  functions to perform MC simulations of the linearly polarized quark and calculated the corresponding number densities in Sec~\ref{SUBSEC_NJL_JET}. The results showed nonzero, but small signal for the helicity dependent DiFF, as depicted in Figs.~\ref{PLOT_D1_GP_PIPL_PIMI}(b) and \ref{PLOT_D1_GP_PIPL_PIMI_ZCUT}(b). The  corresponding analyzing power is small and sharply peaked towards large values of $z$, and the cuts in $z$ often used in experimental setup only moderately enhance  the signal in the medium-$z$ region. For comparison, the analyzing powers of pion Collins fragmentation functions, computed within the same model and plotted in Fig.~8(c) of Ref.~\cite{Matevosyan:2016fwi}, raise quickly and plateaux starting in the small-$z$ region at a value with equal or greater magnitude to the maximum of the analyzing power for the helicity dependent DiFF. These can help to explain the nonobservation of the signal in the existing experimental measurements~\cite{Abdesselam:2015nxn, Vossen:2015znm}. We also explored all the different pion pairs, showing that pairs of positively charged pions have a similar magnitude and opposite sign to the $G_1^\perp$ results expected for $\pi^+\pi^-$ pairs, making them good candidates for future studies.

  Thus far, we discussed the first Fourier cosine moment of $G_1^\perp$ entering into $e^+e^-$ annihilation cross section, defined in \Eq{EQ_G1_MH}. The COMPASS Collaboration~\cite{Sirtl:2017rhi} has measured the SIDIS two hadron production asymmetries $A_{UL}^{\sin(\phi_h-\phi_R)}$ and $A_{UL}^{\sin(2\phi_h-2\phi_R)}$, which contain several of the cosine moments of $G_1^\perp$. The results of our calculations of the first five of these moments for $\pi^+\pi^-$ pairs for $N_L=6$, depicted in Fig.~\ref{PLOT_GP_MOM_PIPL_PIMI_ZCUT}, showed a significant decrease in the size of the signal with the increase of the moment $n$. These results supported the nonobservation of significant asymmetries in COMPASS measurements.
   
   Finally, we validated our MC method in Sec.~\ref{SEC_VALIDATION}. Here we first derived expressions for the integrals of $D_1(z)$ over $z$ for different values of $N_L$, using  combinatoric arguments. We showed, that these analytic results match extremely well with those obtained using the MC method. Further, we derived explicit relations for both $D_1(z)$ and  $G_1^\perp(z)$ for the case of only two hadron production. Again, we found a perfect match to the results obtained using MC simulations. These steps validate our method completely. Moreover, there are two important insights obtained in that section. First, the results in Table~\ref{TABLE_N_PIPI} indicate quite rapid convergence of the results with a $z$-cut, as also seen seen from plots in Fig.~\ref{PLOT_D1_GP_PIPL_PIMI_ZCUT} for the $z$-unintegrated case. Second, \Eq{EQ_GP_NL2} provides an interesting insight into the microscopic mechanism for creating such two-hadron correlations with the longitudinal polarization of  the fragmenting quark. Here the "worm-gear" type splitting function $G_T$ creates a correlation between the transverse spin of the intermediate quark in the hadronization process, which in turn is correlated with the transverse momentum of the second emitted hadron via Collins effect. Thus, it is safe to note, that the Collins effect has a crucial role also in creating  this asymmetry.
   
Future work includes analogous studies of the DiFFs responsible for the two-hadron momentum correlations with the transverse spin of the quark. We performed a detailed study of the dependence of the unpolarized DiFF on both $z$ and $M_h^2$ within the quark-jet model in Ref.~\cite{Matevosyan:2013aka}, where the inclusion of the vector mesons and their strong decays proved crucial to describe the invariant mass spectrum. We also studied what are the relevant contributions of the "primary" emitted pions and koans versus those produced by the vector meson decays to the unpolarized DiFF. The recent experimental and MC studies of the unpolarized DiFFs by  BELLE Collaboration in Ref.~\cite{Seidl:2015lla} strongly support these findings.  The correlations between the  $z$ and $M_h^2$ dependencies were also explored, and it has been shown that different $z$ regions emphasize the contributions of the different resonances in $M_h^2$ spectrum. The main motivation for exploring the $M_h^2$ dependence is to find the signatures of the interference effects between the two hadrons produced in the decays of the resonances, that would generate the polarization-dependent DiFFs. In our work we showed that the interference effects involving just the single hadron production (Collins effect) generates the helicity-dependent DiFF. In this work, we omitted the vector mesons altogether, as the first step in describing the polarized DiFFs within the self-consistent description of the polarized quark hadronization. Thus, we chose not to discuss the $M_h^2$ dependence knowing the obvious lack of the resonant structures there. We leave such detailed analysis to future work, when we consider the vector meson production and their strong decays. Finally, we very roughly mimicked the effects of QCD evolution on DiFFs in this work by using he $(1-z)^4$ ansatz for the input TMD  splitting functions. In future work with a more complete model, we will use QCD evolution for critical comparisons of our results with the experiment.

\section*{ACKNOWLEDGEMENTS}

 The work of H.H.M. and A.W.T. was supported by the Australian Research Council through the ARC Centre of Excellence for Particle Physics at the Terascale (CE110001104), and by an ARC Australian Laureate Fellowship FL0992247 and Discovery Project No. DP151103101, as well as by the University of Adelaide. A.K. was supported by A.I. Alikhanyan National Science Laboratory (YerPhI) Foundation, Yerevan, Armenia.


\bibliographystyle{apsrev4-1}
\bibliography{fragment}

\end{document}